\documentclass[mathleft
]{an}
\usepackage{graphicx}
\usepackage{times}
\begin{document}

\newcommand{\ltae}{\raisebox{-0.6ex}{$\,\stackrel
{\raisebox{-.2ex}{$\textstyle <$}}{\sim}\,$}}
\newcommand{\gtae}{\raisebox{-0.6ex}{$\,\stackrel
{\raisebox{-.2ex}{$\textstyle >$}}{\sim}\,$}}

\Pagespan{1}{}
\Yearpublication{}%
\Yearsubmission{}%
\Month{}%
\Volume{}%
\Issue{}%

\title{Variability of the Fe K line relativistic component in a sample of\\ Seyfert 1 galaxies}

\author{B. De Marco\inst{1,2}\fnmsep\thanks{
  \email{demarco@sissa.it}\newline}
\and  M. Cappi\inst{2}
\and  M. Dadina\inst{2,3}
\and  G. G. C. Palumbo\inst{3}
}
\institute{
S.I.S.S.A./I.S.A.S., via Beirut 4, I-34014, Trieste, Italy
\and 
INAF-IASF, Sezione di Bologna, via Gobetti 101, I-40129 Bologna, Italy
\and 
Dipartimento di Astronomia, Universit\`a degli studi di Bologna, via Ranzani 1, I-40127 Bologna, Italy}

\received{30 Aug 2006}
\accepted{12 Sep 2006}
\publonline{}


\abstract{
We present the analysis of X-ray spectral variability made on a sample of 7 Seyfert 1 bright galaxies, using \emph{XMM-Newton} data. From the ``XMM-Newton Science Archive'' we selected those bright Seyfert 1 showing one or more prominent flares in their 2--10 keV light curves. For each of them we extracted spectra in 3 different time intervals:  before, during and after the flare. We fitted them with a simple power law and then shifted a narrow Gaussian emission  and absorption line template across the 2.5--10 keV data, in order to investigate the presence of line-like features with a confidence level greater than 99\%. Some highly significant features were detected in 3 out of 7 sources studied. In particular, the 3 sources, namely PG 1211+143, NGC 4051 and NGC 3783, showed the presence of a variable emission feature in the 4.5--5.8 keV band, characterized by an increase of its intensity after the flare peak. Because of the observed variability pattern, this feature seems to be ascribable to a reverbered redshifted relativistic component of the Fe K line.}

\maketitle

\section{Introduction}

Radio-quiet Active Galactic Nuclei (AGN),  Seyfert galaxies in particular, show extreme X-ray variability phenomena (Mushotzky, Done \& Pounds 1993). These are characterized by large amplitude flux variations ($\Delta\mathrm{F_{X}/F_{X}}\gtae 1$) on small timescales ($\Delta \mathrm{t< 1\ day}$). Such kind of phenomena are usually indicated with the term of ``flare'' although a rigorous definition of flare doesn't yet exist in literature. The physical mechanisms which produce an AGN X-ray flare are not understood. In particular it is not clear where the flares do originate (e.g. Lightman \& Rybicki 1979, Payne 1980, Poutanen \& Fabian 1999, Malzac \& Jourdain 2000). Despite the present uncertainties on the flares trigger mechanism, such phenomena are useful tools to investigate the innermost regions of AGNs (geometry, chemical composition, black hole mass, etc.). In the context of a disk-corona model (e.g. Haardt \& Maraschi 1991,1993) we expect intense X-ray continuum flux variations to be followed by a disk reflection component response. In particular, if the flare originates close to the central black hole it might be possible to observe a reverbered relativistic Fe K line response (e.g. Fabian et al. 2000). The \emph{XMM-Newton} MCG -6-30-15 observation is the first one in which a reverbered relativistic Fe K line has been detected after a huge hard X-ray flux flare (Ponti et al. 2004).\\
In the following we present a systematic study of X-ray spectral variability in response to a X-ray flare in a sample of bright Seyfert 1 galaxies observed with \emph{XMM-Newton}. The aim is to detect transient emission/absorption features in the Fe K band, produced in response to the flare emission.

\begin{table}
\caption{Sources of the sample: column (1) duration of the flare; column (2) 2--10 keV count rate  variation, with respect to the average count rate, between the maximum and the minimum of the flare.}
\label{tab:sources}
\begin{center}
\begin{tabular}{cccc}\hline
Name & Flare  & $\Delta$t & $\Delta$CR$_{\mathrm{(2-10\ keV)}}$ \\ 
     &         &    (ks)     &                                  \\
     &      &     (1)    &   (2)    \\
\hline
NGC 3783 &  I  &     52 & 0.65\\
         &  II &     80 & 0.81\\
PG 1211+143 & I &    12 & 1.00 \\
TON S 180  & I &      6 & 0.90\\
  &          II&      9 & 0.60 \\
 NGC 4051 &  I &      6 & 1.58\\
  &         II &      8 & 0.88 \\
 I ZW 1 &    I &      6 & 1.00 \\
 PG 1448+273 &I &     4 & 0.61  \\
AKN 564  &  I   &     2 & 1.09\\
\hline
\end{tabular}
\end{center}
\end{table}

\section{Selection of the sample}

\begin{figure}
\includegraphics[width=60mm,height=80mm,angle=270]{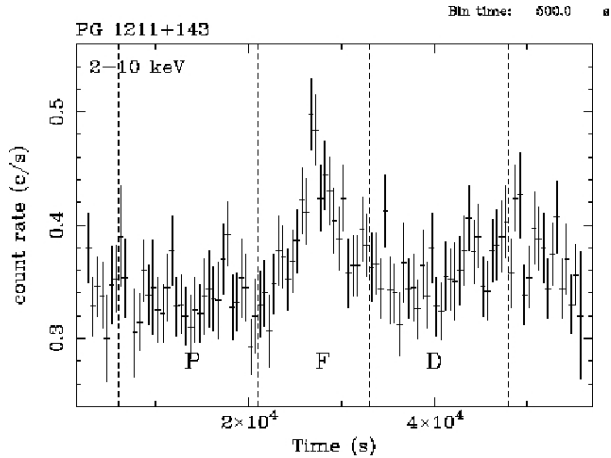}
\caption{PG 1211+143 2--10 keV light curve with time intervals chosen for spectral analysis.}
\label{fig:lightcurvesPG}
\end{figure}

\begin{figure}
\includegraphics[width=60mm,height=80mm,angle=270]{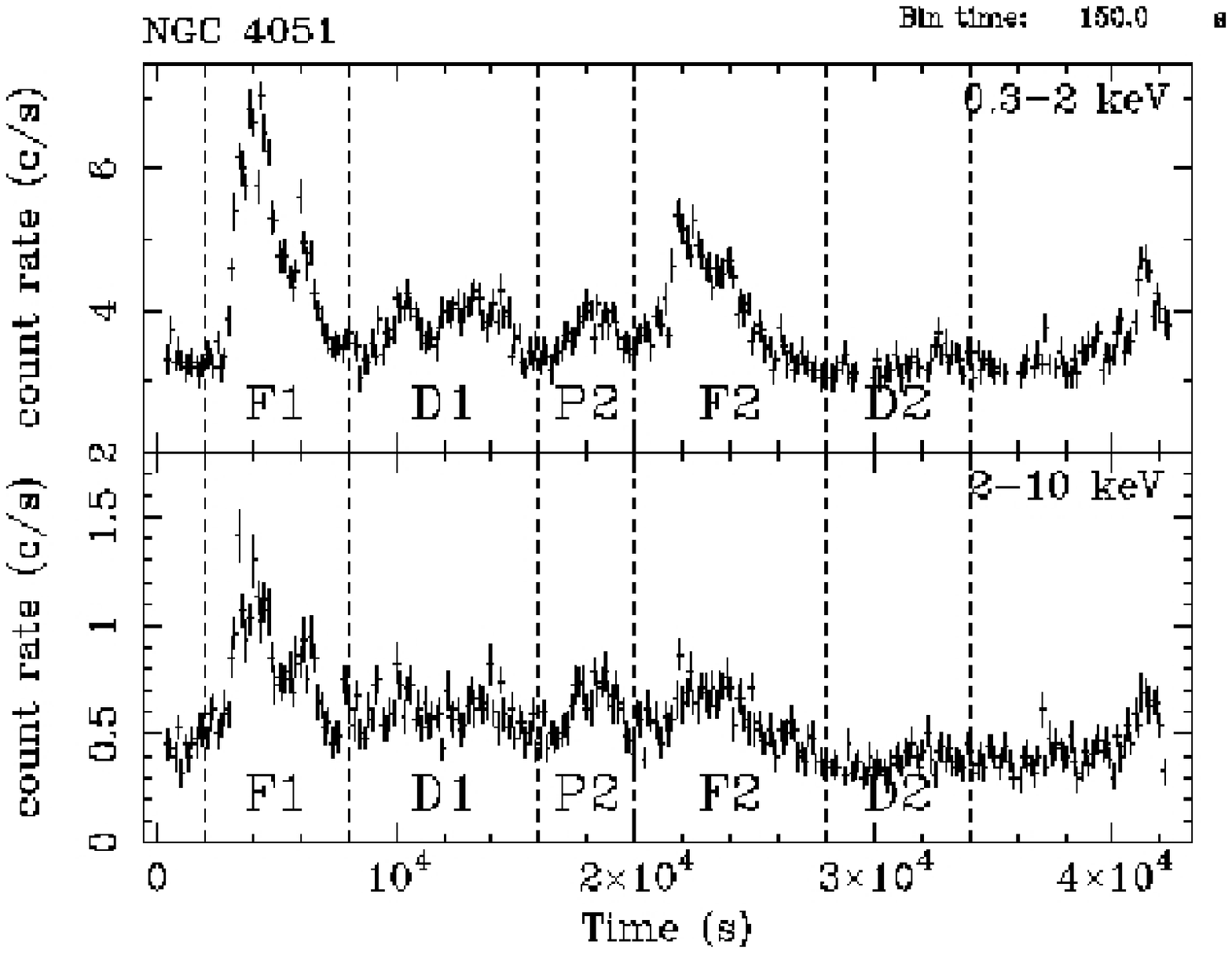}
\caption{NGC 4051 0.3--2 and 2--10 keV light curves with time intervals chosen for spectral analysis.}
\label{fig:lightcurvesNG40}
\end{figure}

\begin{figure}
\includegraphics[width=60mm,height=80mm,angle=270]{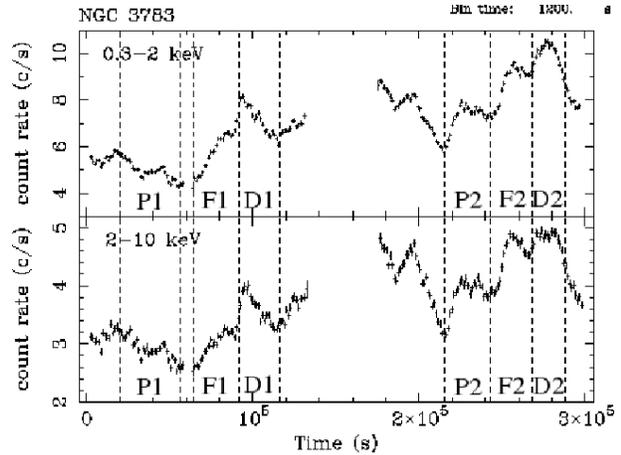}
\caption{NGC 3783 0.3--2 and 2--10 keV light curves with time intervals chosen for spectral analysis.}
\label{fig:lightcurvesNG37}
\end{figure}

The initial sample of sources was obtained cross-correlating the ``XMM-Newton Master Log \& Public Archive'' and the ``V\'{e}ron Quasars and AGNs'' catalog (V\'{e}ron-Cetty, V\'{e}ron \& Gon\c{c}alves 2001) in order to select only Seyfert 1, 1.2 and 1.5 (public data up to 2004 December). The resulting sample was cross-cor\-related with the ``ROSAT All Sky Survey: Bright Sources'' catalog (Voges et al. 1999) in order to associate to each source a mean count rate. We excluded objects with ROSAT count rate $\leq 0.40$ c/s (which, assuming a standard spectrum with $\Gamma=1.7$ and galactic absorption, corresponds to a mean 0.1--2.5 keV flux of $\sim 1.3\times 10^{-11}$ erg cm$^{-2}$ s$^{-1}$); this should avoid the inclusion in the final sample of too weak objects for which a time resolved spectral analysis would be impossible. Also observations lasting less than 10 ks were excluded in order to increase the probability to study the emission properties before, during and after the flares. The final sample contains 67 observations. The 2--10 keV light curves were extracted from the EPIC pn camera data. We looked for those ones in which at least one flare event was present. The flare was defined as a count rate variation between the minimum phase and the peak phase $\geq$ 50\% of the mean count rate registered during the entire observation, i.e. $(\mathrm{CR}_{\mathrm{max}}-\mathrm{CR}_{\mathrm{min}})/\langle \mathrm{CR}\rangle \geq 0.5$. The final sample consists of 7 sources and 10 flares. They are listed in Tab.\ref{tab:sources}; some examples of flares are reported in Fig.\ref{fig:lightcurvesPG}--\ref{fig:lightcurvesNG37}. The MCG -6-30-15 and MKN 766 observations as well as one observation of NGC 4051 were excluded because they are characterized by the superposition of a large number of flares, making hard to identify each of them.


\subsection{Spectral analysis}

\begin{figure}
\begin{center}
\includegraphics[width=60mm,height=75mm,angle=270]{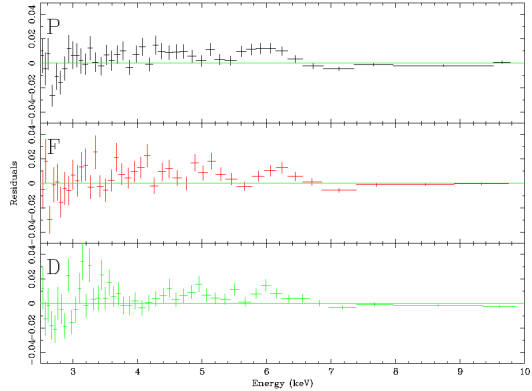}
\caption{PG 1211+143: the 2.5--7.5 keV spectra during time intervals P, F and D. The EPIC pn data are used.}
\label{fig:spectraPG}
\end{center}
\end{figure}

\begin{figure}
\begin{center}
\includegraphics[width=50mm,height=70mm,angle=270]{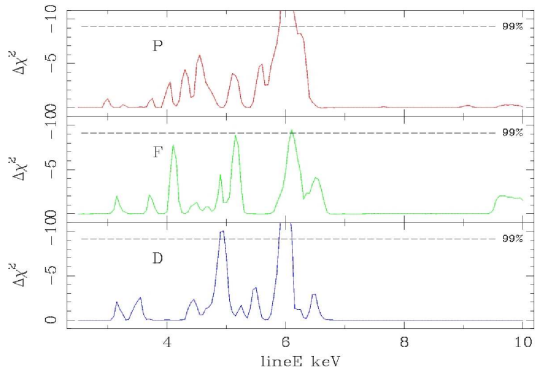}
\caption{PG 1211+143: $\Delta\chi^{2}$ vs. emission gaussian centroid-energy plot in the chosen time intervals.}
\label{fig:examplechi}
\end{center}
\end{figure}

\begin{figure}[!htbp]
\begin{center}
\includegraphics[width=40mm,height=60mm,angle=270]{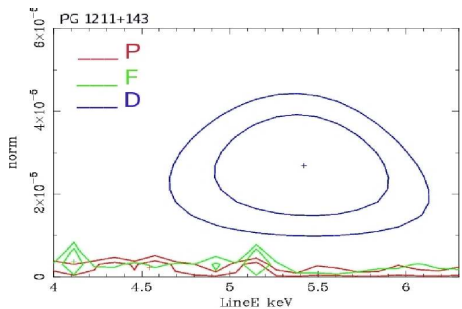}
\caption{PG 1211+143: comparison among the 90\% and 99\% normalization vs. gaussian centroid energy contour plots for the variable feature at $\sim$5.0--5.5 keV in the chosen time intervals. Fe K line intensity during period D was significantly greater than during periods P and F.}
\label{fig:contourplotsPG}
\end{center}
\end{figure}

\begin{figure}[!htbp]
\begin{center}
\includegraphics[width=40mm,height=60mm,angle=270]{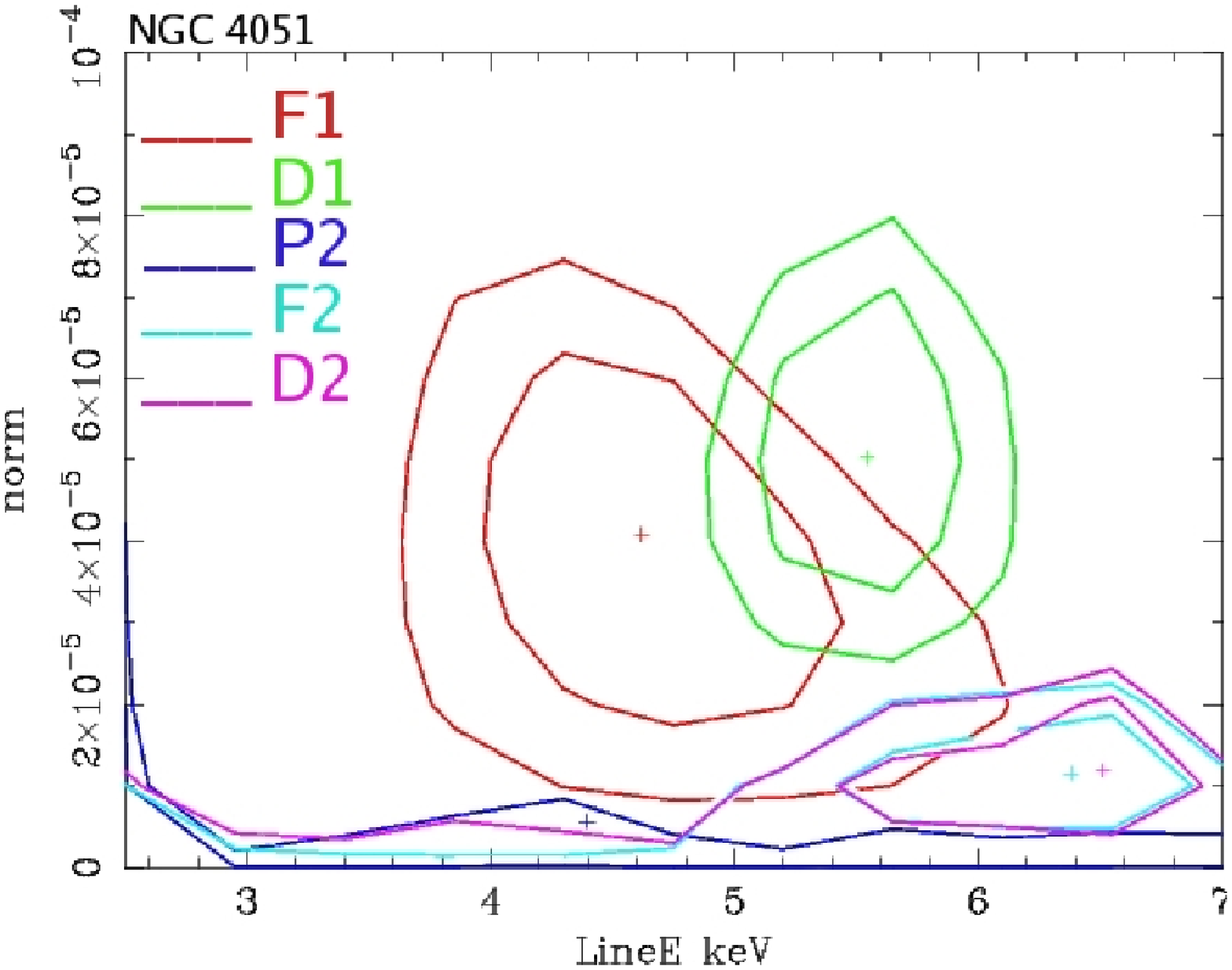}
\caption{NGC 4051: comparison among the 90\% and 99\% normalization vs. gaussian centroid energy contour plots for the variable feature at $\sim$4.5--5.8 keV in the chosen time intervals. Fe K line intensity during periods F1 and D1 was significantly greater than during periods P2, F2 and D2.}
\label{fig:contourplotsNG40}
\end{center}
\end{figure}

\begin{figure}[!htbp]
\begin{center}
\includegraphics[width=40mm,height=60mm,angle=270]{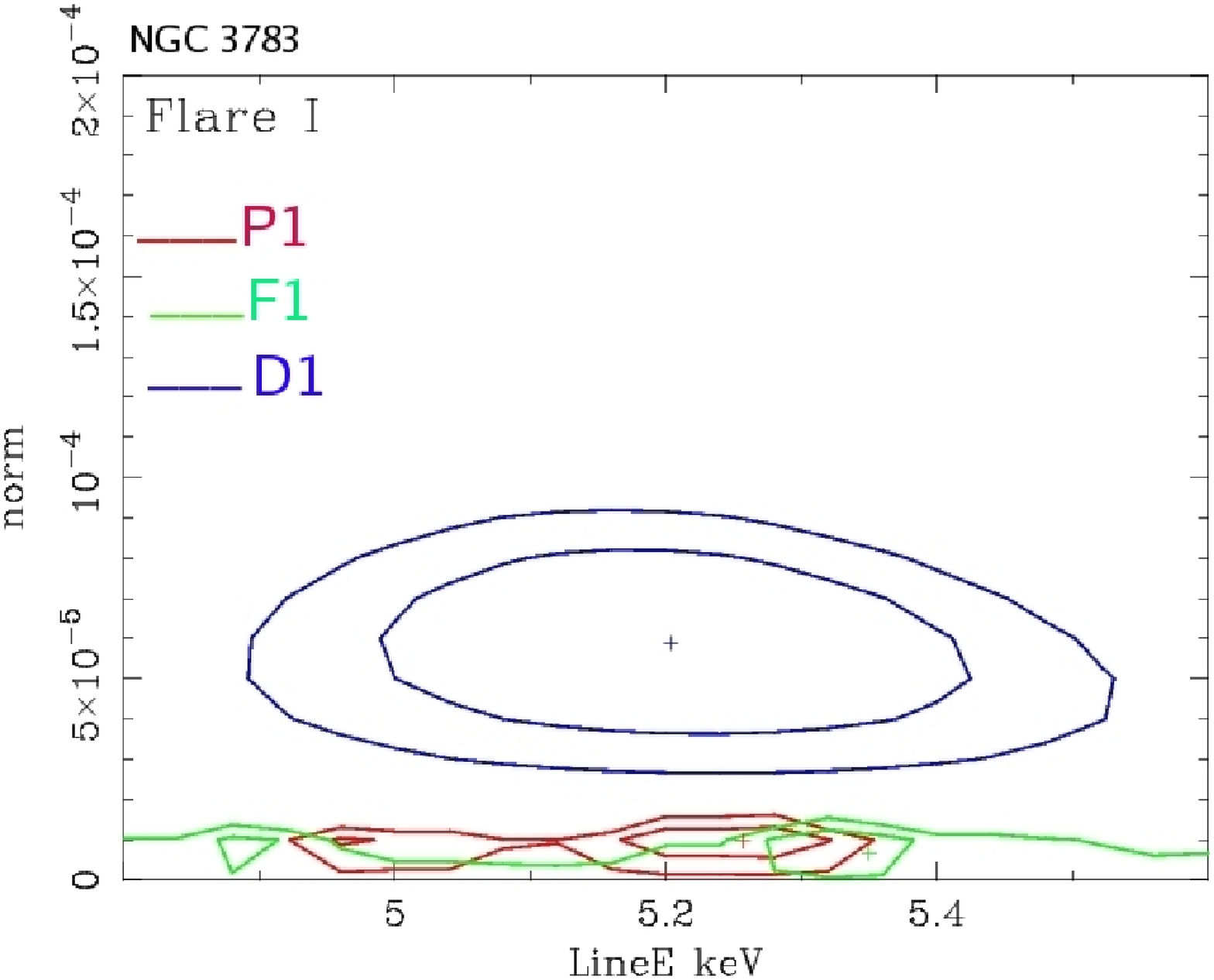}
\includegraphics[width=40mm,height=60mm,angle=270]{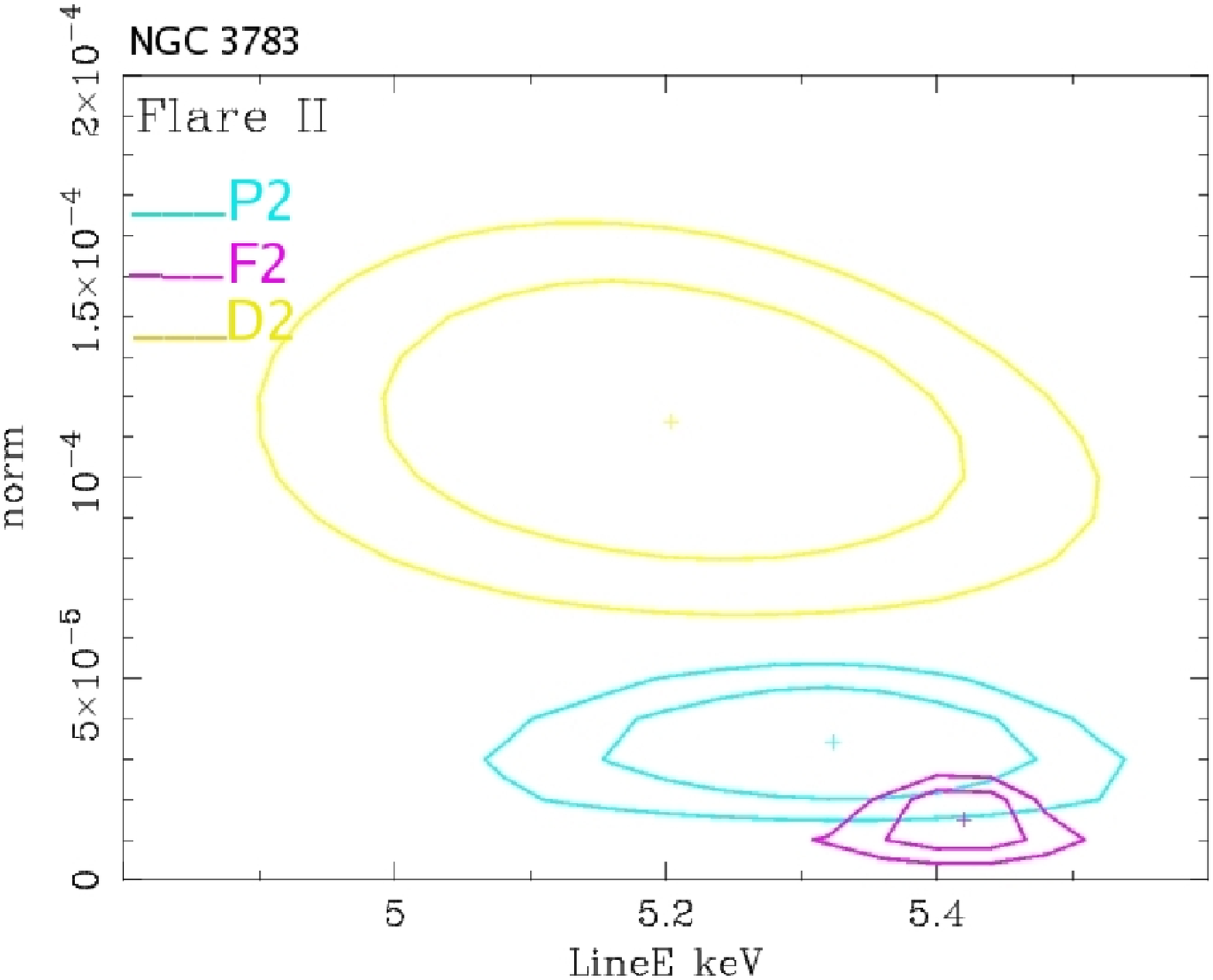}
\caption{NGC 3783: comparison among the 90\% and 99\% normalization vs. gaussian centroid energy contour plots for the variable feature at $\sim$5.1--5.5 keV in the chosen time intervals. Fe K line intensity during periods D1 and D2 was significantly greater than during periods P1, F1, P2 and F2.}
\label{fig:contourplotsNG37}
\end{center}
\end{figure}

The spectra of the 7 sources were extracted in three different time intervals:  before, during and after each flare (hereafter indicated with the letters P, F and D respectively). Because of the limited duration of the observation in the case of NGC 3783 we chose the following time intervals for spectral analysis (see Fig. \ref{fig:lightcurvesNG37}): one time interval characterized by relatively low flux (P1), two time intervals during the first flare (F1 and D1) and three time intervals during the second flare (P2, F2 and D2). Only EPIC pn data were used. The analysis was carried in the 2.5--10 keV band, because in this energy band we expect to observe spectral features associated to the Fe K line, and also to avoid \emph{warm absorber} contamination. In the case of NGC 3783 we considered only the 3.5--10 keV band because in this source the \emph{warm absorber} component extends to higher energies. Each spectrum was initially modeled with a simple power-law. Then a narrow Gaussian emission/absorption line template was added to the model and shifted along the entire energy band at steps of 50 eV. At each step the fit improvement with respect to the simple initial model was computed in terms of $\chi^{2}$ variations. An emission/absorption $\Delta \chi^{2}$ vs gaussian centroid-energy plot was produced in order to search for line-like features with a confidence level greater than 99\% (i.e. $\Delta \chi^{2}>$ 9.21). As an example the 2.5--10 keV spectra of PG 1211+143 are shown in Fig. \ref{fig:spectraPG} and the corresponding $\Delta \chi^{2}$ plots are shown in Fig. \ref{fig:examplechi}. Once significant features were detected, Gaussian lines have been added to the model fixing the width of each Gaussian model to its best fit value. We looked for line-intensity and line-energy variations comparing the Gaussian normalization vs energy parameters contour plots in the different time intervals. Each contour plot was obtained by fixing all the added Gaussians to their best fit values apart from the one for which we wanted to analyze the time variations (see Fig. \ref{fig:contourplotsPG}--\ref{fig:contourplotsNG37}). 


\section{Results and discussion}

Only 3 (PG 1211+143, NGC 4051 and NGC 3783) out of the 7 studied sources show highly significant emission and absorption features. Their light curves are showed in Fig.\ref{fig:lightcurvesPG}--\ref{fig:lightcurvesNG37}. Moreover, these 3 sources display variable emission features in the 4.5--5.8 keV band, the intensity of this component rising after the flare peak (see Fig.\ref{fig:contourplotsPG}--\ref{fig:contourplotsNG37}). This suggests a delayed response with respect to the continuum emission. The variability of the feature is detected with a confidence level greater than 99\%.\\
The hypothesis that the feature might be a curvature of the spectrum produced by variations of the \emph{warm absorber} ionization parameter ($\xi=\mathrm{L_{X}/nR^{2}}$, where n is the density of the absorbing gas and R is its distance from the source) seems to be ruled out by the observed variability pattern: in fact, in this case we should expect a decrease of the spectrum curvature intensity (increase of $\xi$ and, hence, decrease of absorption) after the flare peak. Moreover, the fact that we don't observe contemporary variations of the narrow Fe K$\alpha$ line core component makes implausible the possibility that the feature is due to a reflection {\it continuum} effect. Hence it is most likely ascribable to a reverbered redshifted relativistic component of the Fe k$\alpha$ line, produced in the inner regions of the accretion disk.

\section{Conclusion}

Time-resolved spectral analysis of 7 bright Seyfert 1 has been presented. These sources have been selected for the presence, in their 2.5--10 keV light curve, of at least one prominent flare. For each source and for each flare, we analyzed the data before, during and after the X-ray emission peak. A variable emission feature was observed in 3 up to 7 sources spectra at energies of 4.5--5.8 keV. It shows an increase of intensity after the flare peak at the 99\% confidence level. In our view the feature can be interpreted as a broad highly redshifted component of the Fe K$\alpha$ line produced in the innermost regions of the accretion disk and, hence, subjected to relativistic effects. The variability pattern suggests that it may be produced by the reverberation of the hard X-ray flux continuum. Assuming a disk-corona model, from the measured delays it is possible to give a maximum limit to the distance between the corona and the accretion disk. The obtained values are: 32 r$_{S}$ for PG 1211+143, 800 r$_{S}$ for NGC 4051, 32 r$_{S}$ and 76 r$_{S}$ from the first and the second flare of NGC 3783 respectively.

\acknowledgements
MC, MD and GP thank ASI for financial support under contract I/023/05/0.\\
Based on observations obtained with \emph{XMM-Newton}, an ESA Science mission with instruments and contributions directly funded by ESA Member States and NASA.

\end{document}